# The global consensus on
# the risk management of autonomous driving

Sebastian Krügel[1*]    Matthias Uhl[1]

## Abstract

Every maneuver of a vehicle redistributes risks between road users. While human drivers do this intuitively, autonomous vehicles allow and require deliberative algorithmic risk management. But how should traffic risks be distributed among road users? In a global experimental study in eight countries with different cultural backgrounds and almost 11,000 participants, we compared risk distribution preferences. It turns out that risk preferences in road traffic are strikingly similar between the cultural zones. The vast majority of participants in all countries deviates from a guiding principle of minimizing accident probabilities in favor of weighing up the probability and severity of accidents. At the national level, the consideration of accident probability and severity hardly differs between countries. The social dilemma of autonomous vehicles detected in deterministic crash scenarios disappears in risk assessments of everyday traffic situations in all countries. In no country do cyclists receive a risk bonus that goes beyond their higher vulnerability. In sum, our results suggest that a global consensus on the risk ethics of autonomous driving is easier to establish than on the ethics of crashing.

*Keywords*: self-driving vehicles, risk ethics, intercultural study, social expectations, AI acceptance

[1]University of Hohenheim, Schwerzstraße 46, 70599 Stuttgart, Germany.

*Correspondence to*: sebastian.kruegel@uni-hohenheim.de

*Acknowledgments*: We thank AImotion Bavaria for funding this study.

## Introduction

All driving involves risks of an accident (Goodall, 2016a). Accidents can occur due to technical failure, human error or unforeseeable events (Goodall, 2019). If other road users are around, they are also affected by the risk of a collision. In these ubiquitous situations, every driving maneuver exerts a marginal effect on the risks to other road users (Goodall, 2019; Bonnefon, Shariff & Rahwan, 2019). Often, trade-offs arise in these situations in which the risk of a collision with one road user decreases while the risk of a collision with another increases (Goodall, 2019; Bonnefon, Shariff & Rahwan, 2019; Krügel & Uhl, 2022; Krügel & Uhl, 2024). A car overtaking a cyclist is an example of such a situation (Bonnefon, Shariff & Rahwan, 2019). The greater the lateral distance to the cyclist, the greater the risk of a collision with a road user on the other side of the road. The smaller the distance to the cyclist, the greater the risk of a collision with that cyclist. Lateral lane positioning, in general, is a typical example of these risk considerations (Goodall, 2016a; Goodall, 2019). Various braking strategies for avoiding a collision with a leading vehicle while simultaneously risking a collision with a trailing vehicle is another typical example of such risk considerations (Goodall, 2019).

The distribution of risks to other road users depending on one's own driving behavior has clear ethical implications. When risks are redistributed between different road users, dilemma situations arise in a risk-ethical context (Nyholm & Smids, 2016; Geisslinger et al., 2021; Krügel & Uhl, 2022; Krügel & Uhl, 2024). In contrast to the unrealistic moral dilemmas in trolley-like problems (Foot, 1967) of unavoidable accident scenarios, these dilemmas are part of everyday traffic situations (Goodall, 2019; Krügel & Uhl, 2024). While human road users solve these risk-ethical dilemmas intuitively, AVs must do so on the basis of a carefully planned risk management (Goodall, 2016a, b). What factors should play a role in this risk management and who decides this?

Much of the engineering literature on AVs is exclusively concerned with minimizing the probability of accidents (see, e.g., Reichardt & Shick, 1994; Gehrig & Stein, 2007; Erlien, Fujita & Gerdes, 2013; Wolf & Burdick, 2008; Keller et al., 2014; Funke et al. 2015; Wachenfeld et al., 2016; Gerdes & Thornton, 2016; Thornton et al., 2016; Claussmann et al., 2020). Collision avoidance is often seen as a "deontological rule" (Thornton et al., 2016) that takes "the highest priority of the automated vehicle" (Thornton et al., 2016). Other factors, such as accident severity, tend to be ignored in the engineering literature on the risk management of AVs. Some automotive companies seem to be going a bit further. In some of its patents, *Google Inc.* and later *Waymo LLC*, for instance, discuss the possibility of taking other factors into account in the risk management of AVs, such as accident severity, as well as the type, size and vulnerability of the other road users, in addition to the probability of an accident (see Dolgov & Urmson, 2014; Teller & Lombrozo, 2014, 2015, 2019). In turn, the ethical literature on AVs largely focuses exclusively on accident severity and the distribution of damages (see, e.g., Bonnefon, Shariff & Rahwan, 2016; Awad et al., 2018; Faulhaber et al., 2019; Frank et al., 2019; Huang, Greene & Bazerman, 2019; Bigman & Gray, 2020; Morita & Managi, 2020), and mostly disregards accident probabilities altogether.

Apparently, the prevailing view of the relevant factors in the risk management of AVs depends heavily on the discipline. This is unsatisfactory, of course. Another approach is surveying the public's views on AVs' risk management. First, it is important to know what the public expects of AVs' behavior in road traffic, not least in order to address a possible discrepancy between public's expectations and actual behavior of AVs in a public discourse (see, e.g., Bonnefon, Shariff & Rahwan, 2016; Adnan et al., 2018; Awad et al., 2018; Cunningham et al., 2019; Krügel & Uhl, 2022). Second, the public constitutes the group of those affected by AVs – they bear the risks arising from AVs' behavior in road traffic. The public should therefore also have a say in how AVs should ideally behave on the road (Bonnefon, Shariff & Rahwan, 2019; Krügel & Uhl, 2024). It should be emphasized here that desired and technically feasible driving behavior are two completely different categories. The desired behavior describes the first-best solution of AVs' behavior. It is then the task of engineering to find out whether and to what extent this solution is technically feasible.

To the best of our knowledge, Krügel & Uhl (2024) is the only study that explicitly surveys the public's views on the risk distribution of AVs in everyday road traffic. They investigated whether the public considers minimizing the probability of accidents in everyday traffic situations as a guiding principle as it is done in large parts of the engineering literature. In a representative survey in Germany, participants in their study were asked to adjust the lateral lane positioning of a self-driving car until the desired distances between other road users were achieved. Along with this lane positioning adjustment, accident probabilities also changed, so that the risks between road users were redistributed. The most important results of the study were, firstly, that the participants not only took into account the probability of accidents when positioning the AV, but also the severity of the accident. Thus, collision avoidance as the guiding principle in road traffic was not in line with the views of the participants in their study. Secondly, the study participants indicated a willingness to take risks themselves for the benefit of other road users. The "social dilemma of AVs" (Bonnefon, Shariff & Rahwan, 2016), which arises in unavoidable accident situations, is apparently less pronounced when it comes to the risk management of AVs in everyday road traffic.

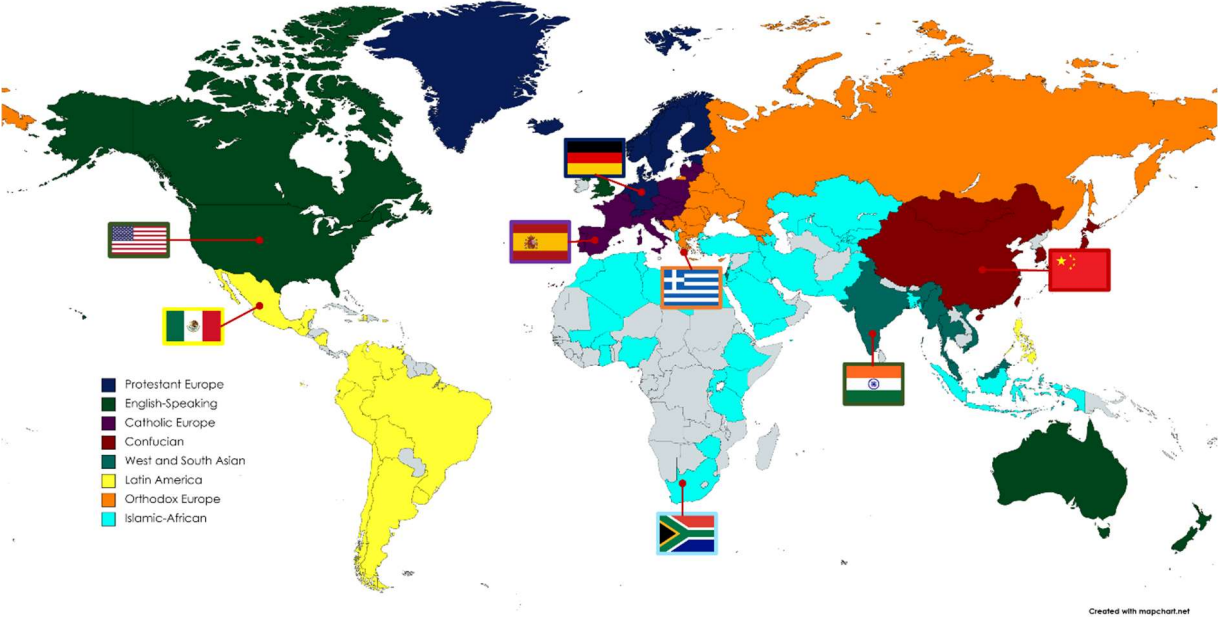

**Figure 1. Countries of our study distributed around the world and the Inglehart-Welzel Cultural Map.** The figure shows the eight countries included in our study. We chose one country from each of the eight clusters that are based on the 2023 version of the Inglehart-Welzel Cultural Map of the World.

The present paper builds on the study by Krügel & Uhl (2024) and expands on it in one important dimension. We examine how people's risk distributions in road traffic vary across different countries and cultures. Based on the Inglehart-Welzel Cultural Map of the World from 2023 (World Values Survey 7, 2023), we selected one country from each of the eight cultural clusters for our study (see also *Figure 1*): China (Confucian), Germany (Protestant Europe), Greece (Orthodox Europe), India (African-Islamic), Mexico (Latin America), South Africa (West & South Asia), Spain (Catholic Europe) and USA (English-Speaking). In each of these countries, we conducted an online survey in the local language. We surveyed 1,372 people in each country, for a grand total of 10,976 participants. With the help of a survey service provider that handled the recruitment of participants, we ensured that we had a balanced, online-representative sample of participants aged 18 and older in all countries.

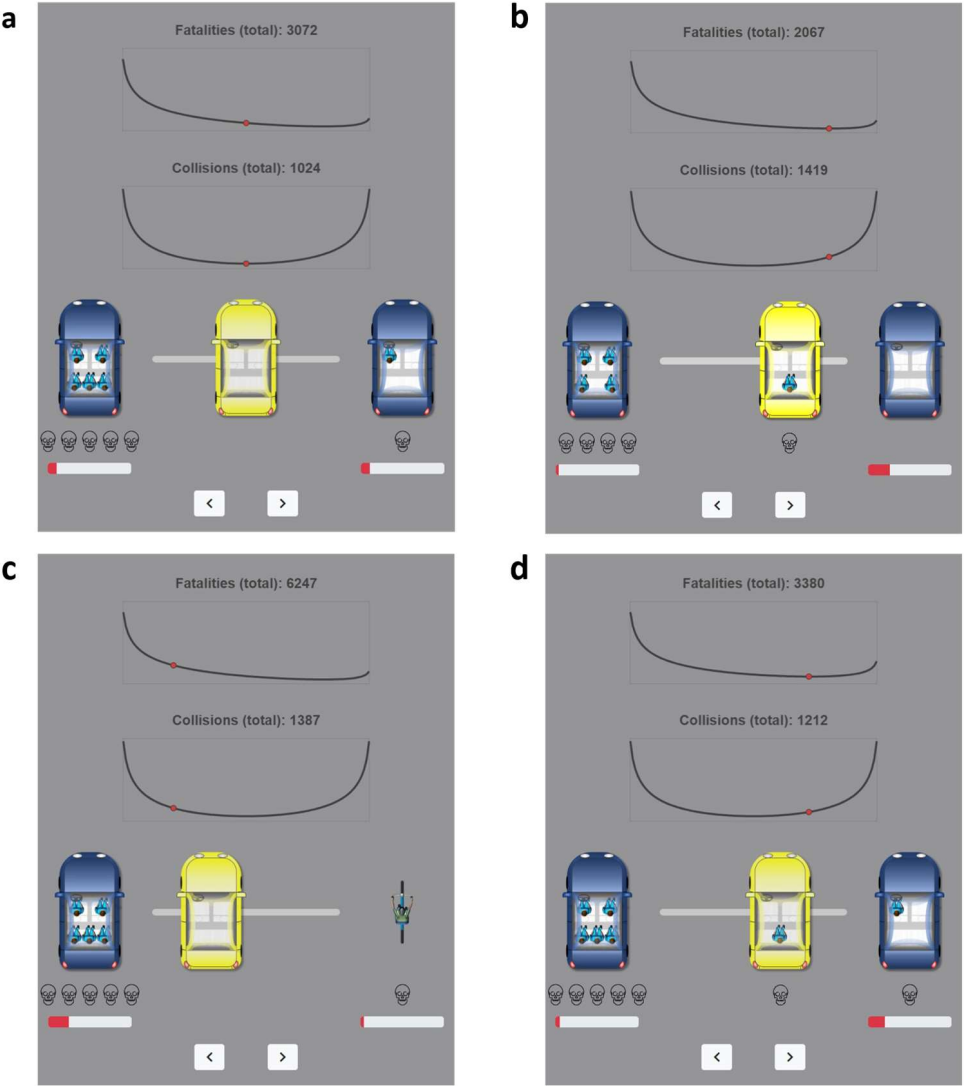

**Figure 2. Graphical interface for eliciting preferences about risk allocation in road traffic.**
The figure shows four traffic situations of a self-driving car (i.e., the yellow AV in the middle of each setting). The AV could be moved to the left and right along the light gray line in 99 increments. When the AV was moved, the red dots on the graphs moved accordingly, the statistics above the graphs adjusted and the red bars below the vehicles on the left and right side of the road increased or decreased in size. The order of the graphs and whether the majority of passengers in the blue cars appeared on the left or right side of the road was determined at random for each participant. If a cyclist was part of the traffic situation, the cyclist was always shown on the right side of the road.

To be able to investigate globally how much people consider accident probability and severity in risk distributions in road traffic, we modified the methodology of Krügel & Uhl (2024). We provided the participants with two traffic statistics for each situation: the total number of accidents and the total number of fatalities in one million such traffic situations (see *Figure 2*). Depending on the chosen lateral lane positioning of the AV, these statistics changed and the red dots moved accordingly along the corresponding graph of the function. It should be emphasized that it was not possible to minimize both traffic statistics at the same time. The minima of both statistics were linked to different lane positionings of the AV. This allowed us to examine how participants weigh the two factors in their risk distribution. The total number of accidents per one million such traffic situations corresponds to the probability of an accident. The total number of fatalities represents the severity of an accident.

In our study, we used several everyday traffic situations which always involved weighing the probability of an accident against the severity of the accident. Each participant was asked to distribute the traffic risks in one randomly selected situation. In all situations, a (yellow) AV was driving between two other vehicles (see *Figure 2*). The lateral lane positioning of the (yellow) AV could be adjusted by the participants at their own discretion by moving the AV further to the left or right. The initial position of the AV was chosen at random for each participant. In total, lane positioning of the AV could be customized in 99 steps. With each new positioning, the two traffic statistics also changed. The accident probability increased exponentially the closer the AV was positioned to another vehicle. The overall accident probability was lowest when the AV was positioned exactly in the middle of the lane. The total number of fatalities depended on the accident probability and the number of road users involved. If the AV was moved away from the middle of the lane towards the vehicle with fewer passengers, the total number of fatalities decreased at first. As the distance to this vehicle narrowed, the accident probability increased to such an extent that the total number of fatalities rose again with each further reduction in the distance. Thus, there was a lane positioning of the AV that minimized the total number of fatalities in the corresponding traffic situation, and this positioning was away from the middle of the lane.

*Figure 2* shows all variants of different passenger constellations in the traffic situations used in our study. When the (yellow) AV was driving between two cars (see *Figure 2a, b, d*), the majority of passengers were randomly displayed either on the left or right side of the road. When the AV was driving between a single cyclist and a car with five passengers (see *Figure 2c*), the cyclist was always shown on the right side of the road. In some situations, the (yellow) AV was driving without passengers (see *Figure 2a, c*). In these situations, the participants in our study were not part of the traffic situation. In other situations, the (yellow) AV was driving with one passenger (see *Figure 2b, d*). In these situations, we made it clear to the participants in our study that they were part of the traffic situation by being the passenger of the (yellow) AV (see *Figure 6* in section *Methods*). If the AV with the participants as a passenger was driving between two cars with five passengers on one side and one passenger on the other side of the road (see *Figure 2d*), the traffic statistics concerning the total number of fatalities were altered. This was due to the fact that the passenger in the AV represented an additional casualty. In order to account for this, we reduced the number of passengers in the cars on the left and right of the road by one person in two traffic situations in which the AV was driving with a passenger (see *Figure 2b*). In this way, the traffic statistics could be kept constant between situations to examine how lateral lane positioning of the AV changes when the participants of our study were part of the traffic situation.

## Results

*Mean driving positions per country*

We first examine the mean positioning in each country in situations in which the AV was traveling between two cars. We compare situations in which an empty AV was traveling between two cars with one or five passengers with situations in which an AV with the participants as a passenger was traveling between two cars with either no or four passengers. The side of the road with the majority of passengers was randomly selected for each participant. The positioning of the AV by the participants was converted into a variable with integer values between *-49* and *+49* for the following analysis. Values of zero reflect traveling at the lane's center; positive (negative) values reflect a deviation from the lane's center in the direction of the car with the smaller (larger) number of passengers. Minimizing the number of accidents could be achieved by positioning the AV at the lane's center (i.e., at *0*); minimizing the number of fatalities could be achieved by deviating from the lane's center and occurred at a value of *+30*. In situations in which an AV with the participants as a passenger was traveling between two cars with one or five passengers, the minimization of the number of fatalities occurred at a position closer to the lane's center (i.e., at *+22*). Indeed, participants in these situations positioned the AV significantly closer to the lane's center, on average, than in situations in which they were traveling in an AV between cars with no or four passengers (*8.99* vs. *13.46*, *t(6224.8) = -7.36*, *p < 0.001*). To keep the minima of the traffic statistics constant across situations, those in which an AV with the participants as a passenger was traveling between two cars with one or five passengers were therefore not considered in the following analysis.

*Figure 3* shows that the mean positionings of the AV in all countries deviate from the lane's center in the direction of the car with the smaller number of passengers. This is the case both when the AV is empty and when the participants are part of the traffic situation as a passenger of the AV. The deviations from the lane's center are statistically significant in all countries (see columns 3 to 5 and 7 to 9 in *Table 1*). On average, the participants' views on the driving behavior of AVs thus deviate from a guiding principle of pure accident avoidance. When it comes to risk management in road traffic, participants in all countries are, on average, of the opinion that the severity of accidents should not be ignored. More severe accidents should be avoided as far as possible by shifting the accident probabilities accordingly, even if this disproportionately increases the probability of less serious accidents.

Surprisingly, the mean positionings between situations in which the AV is empty and in which the participants are a passenger of the AV are almost identical in each country (see red and blue bars in *Figure 3* in comparison). We do not find a statistically significant difference between the means in these two situations in any of the countries (see columns 10 to 12 in *Table 1*). This is notable, as previous studies based on deterministic trolley problems have found that people prefer AVs that protect them as passengers at all costs (Bonnefon, Shariff & Rahwan, 2016). We thus confirm the results of a study in Germany, which finds in a risk context that people are willing to accept minimal risks if this makes more severe accidents less likely for other road users (see Krügel & Uhl, 2024). We considerably extend these results and find no evidence for the existence of a "social dilemma of AVs" (Bonnefon, Shariff & Rahwan, 2016) in any of our eight studied countries when situations with risk are considered.

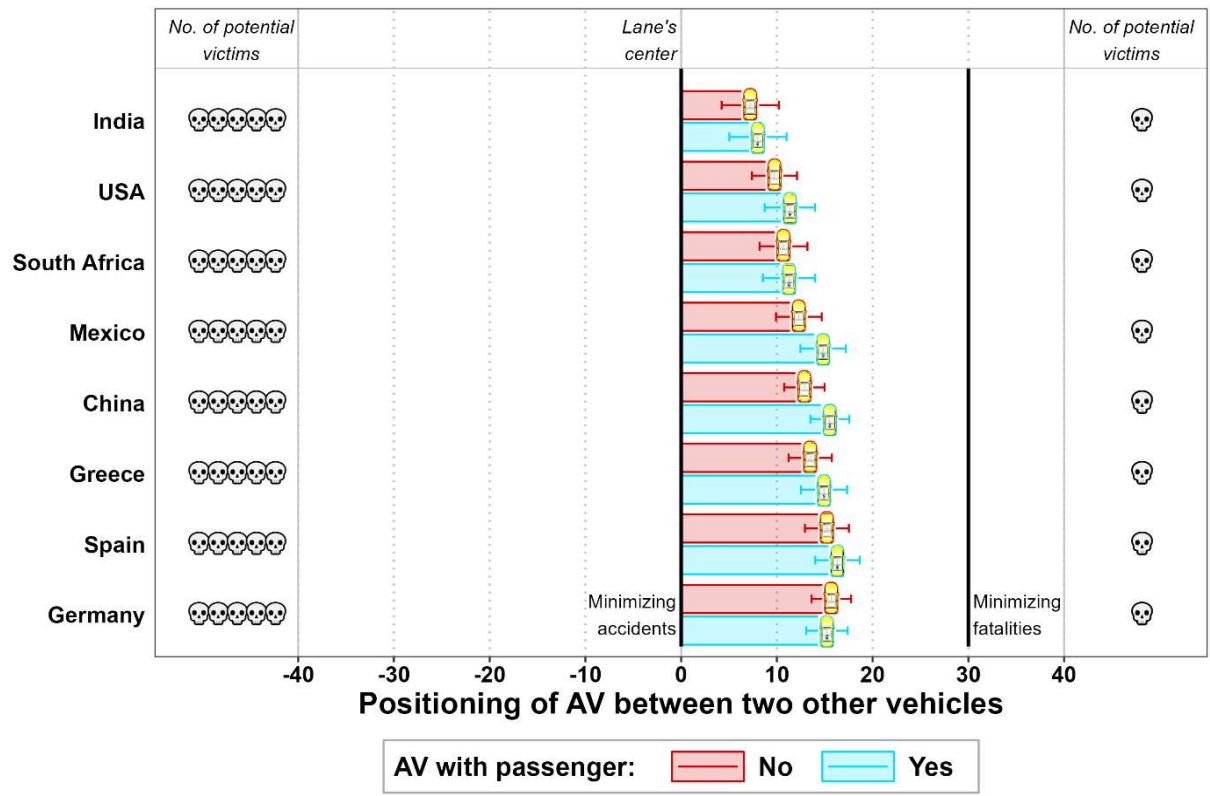

**Figure 3. Average positioning of the AV between road users of the same type.**
The figure shows means and 95% confidence intervals of participants' positioning of the (yellow) AV in each country. The side of the road with the majority of passengers in the (blue) cars was randomly selected for each participant. The bars therefore show mean deviations from the lane's center towards the car with fewer passengers, regardless on which side of the road this car was shown. Red shaded means and bars show the results for the cases when the participants themselves were not part of the traffic situation (i.e., the AV was empty). Blue shaded means and bars show the results for the cases when the participants were part of the traffic situation as a passenger of the AV.

| | AV without passenger | | | | AV with passenger | | | | AV with vs. without passenger | | |
|---|---|---|---|---|---|---|---|---|---|---|---|
| | *mean* | *t* | *df* | *p* | *mean* | *t* | *df* | *p* | *t* | *df* | *p* |
| Germany | 15.68 | 14.88 | 406 | <.001 | 15.23 | 13.81 | 381 | <.001 | 0.30 | 782.3 | .768 |
| Spain | 15.23 | 13.03 | 397 | <.001 | 16.33 | 13.81 | 392 | <.001 | -0.66 | 788.7 | .507 |
| Greece | 13.48 | 11.79 | 390 | <.001 | 14.93 | 12.17 | 397 | <.001 | -0.86 | 784.1 | .389 |
| China | 12.88 | 11.99 | 385 | <.001 | 15.53 | 15.03 | 401 | <.001 | -1.78 | 783.3 | .075 |
| Mexico | 12.30 | 10.12 | 392 | <.001 | 14.83 | 12.30 | 386 | <.001 | -1.48 | 778.0 | .139 |
| South Africa | 10.69 | 8.41 | 398 | <.001 | 11.27 | 8.12 | 384 | <.001 | -0.32 | 773.6 | .759 |
| USA | 9.75 | 8.14 | 399 | <.001 | 11.36 | 8.49 | 382 | <.001 | -0.89 | 767.8 | .372 |
| India | 7.22 | 4.74 | 391 | <.001 | 8.02 | 5.27 | 388 | <.001 | -0.37 | 779.0 | .709 |

**Table 1**: **Participants' positioning of the AV per country**
The table shows mean values of participants' positioning of the (yellow) AV in each country when there was a (blue) car on either side of the road. Positive values indicate a mean deviation from the lane's center towards the car with fewer passengers. Columns 3 to 5 ("AV without passenger") and 7 to 9 ("AV with passenger") show the results of two-sided *t*-tests per country to determine whether the respective mean positionings deviate significantly from the lane's center. Columns 10 to 12 show the results of two-sided *t*-tests per country to determine whether the mean positionings differs significantly when the participants were passengers in the (yellow) AV or not.

We secondly examine whether participants' allocation of risk in road traffic depends on the type of road users involved. For this purpose, we compare situations in which an empty AV was traveling between a car with five passengers on the left and a car with one passenger on the right side of the road with situations in which there was one cyclist instead of a car with one passenger on the right side of the road. The traffic statistics shown to the participants were identical in these situations. This means that an accident on the right side of the road was fatal for the person involved in our study, regardless of whether this person was in a car or on a bicycle. It was not our intention to investigate whether people think that cyclists should be given a greater safety distance because they are more vulnerable. What we wanted to study was whether people think that cyclists should be given a greater safety distance than car users even in the case of identical risks for both road users. It would be ethically justifiable, for example, that cyclists are granted a risk bonus in road traffic because they themselves pose a lower risk to other road users than car users do.

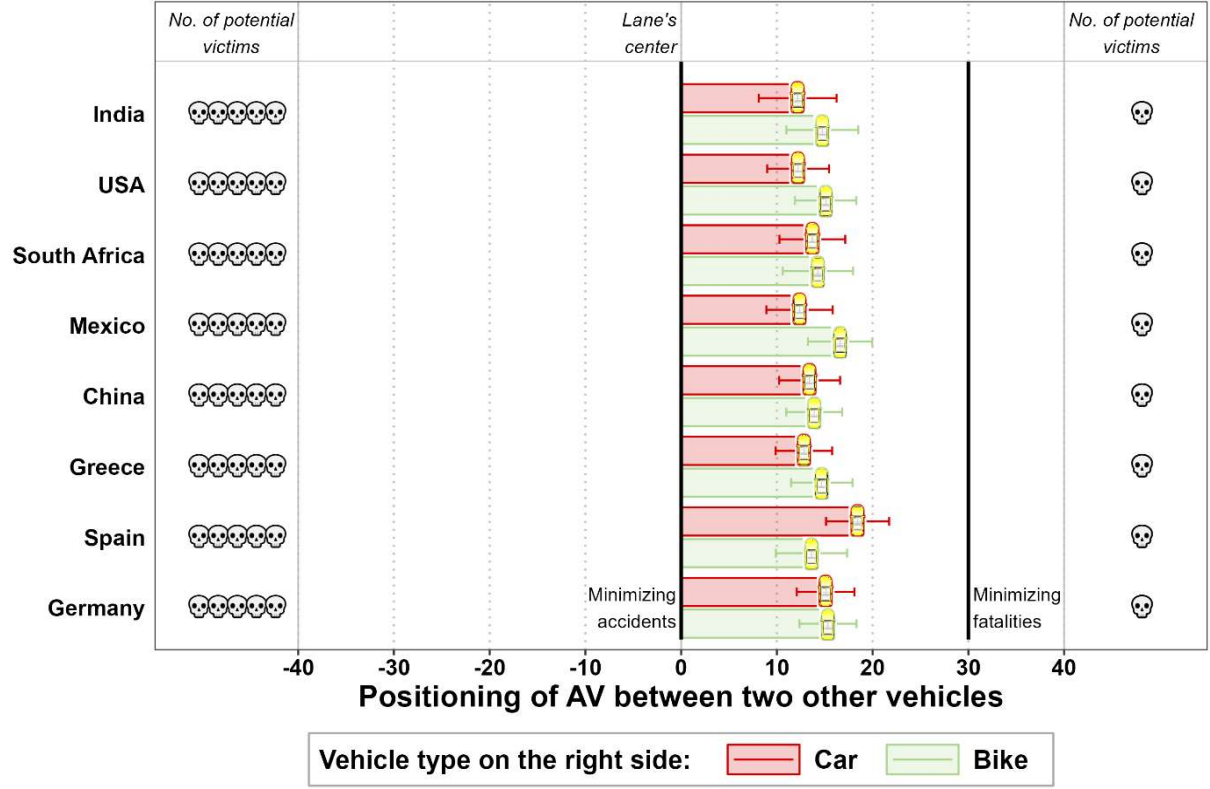

**Figure 4. Average positioning of the AV between road users of different types.**
The figure shows means and 95% confidence intervals of participants' positioning of the (yellow) AV in each country. The data contains only cases when there was a car with five passengers on the left side of the road and a car with one passenger or one cyclist on the right side of the road. The (yellow) AV was always empty. The bars show mean deviations of the AV from the lane's center towards the right side of the road. Red shaded means and bars show the results for the cases when there was a car with one passenger on the right side of the road. Green shaded means and bars show the results for the cases when there was one cyclist on the right side of the road.

As can be seen in *Figure 4*, this is not the case. On average, the positionings of the AV between situations in which there was a car with one passenger and in which there was a cyclist on the right side of the road are very similar in each country (see red and green bars in *Figure 4* in comparison). We do not find a statistically significant difference between the means in these two situations in any of the countries (see columns 10 to 12 in *Table 2*). It appears that the participants in our study did not distinguish between the types of road users involved in the allocation of risk when the risk of a fatal accident was held constant for both types by design. Furthermore, the results are similar to those in *Figure 3*. The mean positionings of the AV deviate significantly from the lane's center in the direction of the single victim in all countries, regardless of whether this victim is a passenger of a car (see columns 3 to 5 in *Table 2*) or a cyclist (see columns 7 to 9 in *Table 2*).

|  | Car with one passenger on the right side | | | | One cyclist on the right side | | | | Car vs. cyclist on the right side | | |
| --- | --- | --- | --- | --- | --- | --- | --- | --- | --- | --- | --- |
|  | *mean* | *t* | *df* | *p* | *mean* | *t* | *df* | *p* | *t* | *df* | *p* |
| Germany | 15.09 | 9.87 | 198 | <.001 | 15.33 | 10.15 | 188 | <.001 | -.113 | 385.0 | .910 |
| Spain | 18.43 | 11.03 | 189 | <.001 | 13.62 | 7.23 | 187 | <.001 | 1.91 | 370.2 | .057 |
| Greece | 12.83 | 8.58 | 201 | <.001 | 14.69 | 9.01 | 190 | <.001 | -.841 | 386.0 | .401 |
| China | 13.41 | 8.31 | 195 | <.001 | 13.90 | 9.37 | 195 | <.001 | -.223 | 387.3 | .823 |
| Mexico | 12.37 | 7.04 | 201 | <.001 | 16.61 | 9.76 | 195 | <.001 | -1.73 | 395.9 | .084 |
| South Africa | 13.71 | 7.87 | 204 | <.001 | 14.28 | 7.67 | 192 | <.001 | -.224 | 392.3 | .823 |
| USA | 12.21 | 7.46 | 198 | <.001 | 15.09 | 9.29 | 193 | <.001 | -1.25 | 391.0 | .212 |
| India | 12.17 | 5.90 | 185 | <.001 | 14.75 | 7.74 | 188 | <.001 | -.919 | 370.1 | .359 |

**Table 2**: **Participants' positioning of the AV per country**
The table shows mean values of participants' positioning of the (yellow) AV in each country when there was a (blue) car with one passenger or one cyclist on the right side of the road. The (yellow) AV was always empty. Positive values indicate a mean deviation from the lane's center towards the right side of the road. Columns 3 to 5 ("Car with one passenger…") and 7 to 9 ("One cyclist…") show the results of two-sided *t*-tests per country to determine whether the respective mean positionings deviate significantly from the lane's center. Columns 10 to 12 show the results of two-sided *t*-tests per country to determine whether the mean positionings differs significantly when there was a car with one passenger or one cyclist on the right side of the road.

*Mean driving positions across countries*

A similar pattern of AV positioning emerged in all countries. The mean positionings deviated significantly from the lane's center towards the smaller number of possible accident victims, regardless of whether the participants themselves were part of the traffic situation or not, and whether a car with a passenger or a cyclist was traveling next to the AV. Thus, on average, the participants in all countries took the severity of accidents into account when deciding about AV positioning. The question we will now investigate is whether the extent to which accident severity is taken into account differs between countries. In situations in which a cyclist was traveling on the right side of the road, this is not the case. Here we do not find statistically significant differences between the means of the countries ($F(7) = 0.30$, $p = 0.95$). If the AV was traveling between two cars, the means of the countries differed significantly, both when the AV was traveling without passengers ($F(7) = 5.51$, $p < 0.001$) and when the AV was traveling with the participants as a passenger ($F(7) = 5.29$, $p < 0.001$).

*Table 3* shows all significant country comparisons in these situations based on pairwise post-hoc tests. The *p*-values were adjusted based on the Tukey method (TukeyHSD, Tukey, 1949). Applying Bonferroni and Holm adjustments to the *p*-values (Holm, 1979) leads to same results. As can be seen in *Table 3*, the mean positioning of participants from the USA differs significantly to participants from Germany and Spain in situations in which the AV was empty. Compared to these two countries, the mean positioning in the USA deviates less from the lane's center towards the car with the smaller number of passengers in these situations. All other significant country comparisons revolve exclusively around India. The mean positionings in India also deviate less from the lane's center towards the car with the smaller number of passengers compared to some other countries in situations in which the AV was empty and in which the participants were a passenger of the AV. Accident severity in India is taken into account (see the results in *Table 1*), but apparently to a lesser extent than in some other countries (see the results in *Table 3*).

|  | No Passenger in AV | | | | Passenger in AV | |
|---|---|---|---|---|---|---|
|  | *India* | | *USA* | | *India* | |
|  | Δposition | p | Δposition | p | Δposition | p |
| Germany | -8.46 | <.001 | -5.93 | 0.012 | -7.21 | .001 |
| Spain | -8.01 | <.001 | -5.47 | 0.030 | -8.31 | <.001 |
| Greece | -6.26 | .007 | --- | --- | -6.91 | .002 |
| China | -5.66 | .024 | --- | --- | -7.51 | <.001 |
| Mexico | --- | --- | --- | --- | -6.81 | .003 |

**Table 3. Pairwise post-hoc tests in situations when there was a car on either side of the road.** The table shows all significant differences of mean positionings of the (yellow) AV in pairwise post-hoc tests between countries. *p*-values were adjusted based on the Tukey method (TukeyHSD, Tukey, 1949). *Δposition* indicate the difference of the mean positioning from the respective country in column 1. Negative values indicate a smaller deviation from the lane's center (i.e., the mean positioning was closer to the lane's center). All comparisons not listed were not significant in pairwise post-hoc tests.

*Distributions of driving positions across countries*

Once again, the pattern of mean AV positioning is very similar across all countries in our study. On average, the participants in all countries take into account the severity of possible accidents and adjust lateral lane positioning of the AV accordingly. Even the extent of this adjustment is very similar in all countries. The only country that differs slightly from the others in the extent of lateral lane adjustment is India. In the final part of the result section, we will take a closer look at the distributions of AV positioning in each country and examine whether these differ substantially between countries.

*Figure 5* shows the distributions of AV positioning in each country (Panels a to c) and in all countries together (Panel d) in a series of ridgeline plots. Panel (a) shows the country-specific distributions in situations in which the AV was empty, panel (b) in those in which the participants were a passenger of the AV, and panel (c) in those in which a cyclist was traveling on the right side of the road. The histograms show the density of AV positioning, with each bar comprising three adjacent AV positions for better visualization. Bars at *0*, for instance, include positionings of the AV exactly at the lane's center as well as one deviation to the left and one to the right.

As can be seen in *Figure 5*, the pattern of the distributions is very similar in all countries. Irrespective of the specific traffic situation, there is a cluster in each country at the position where the number of accidents is minimized (i.e., at the lane's center), at the position where the number of fatalities is minimized (i.e., at *+30*), and at both tails of the distribution (at *-49* and *+49*), where there is a complete shift of traffic risks from road users on one side to road users on the other side. Except for India, however, the cluster at the tail of the distribution where there is a complete shift of risks from the smaller to the larger group of road users (i.e., at *-49*) is rather small. In addition to the four peaks in the distributions, there is an additional cluster of AV positionings between the two minima of the traffic statistics (i.e., between the lane's center and *+30*). These positionings are due to participants who apparently do not favor a pure minimization of the number of accidents nor a pure minimization of the number of fatalities and instead seem to prefer a mixture of both.

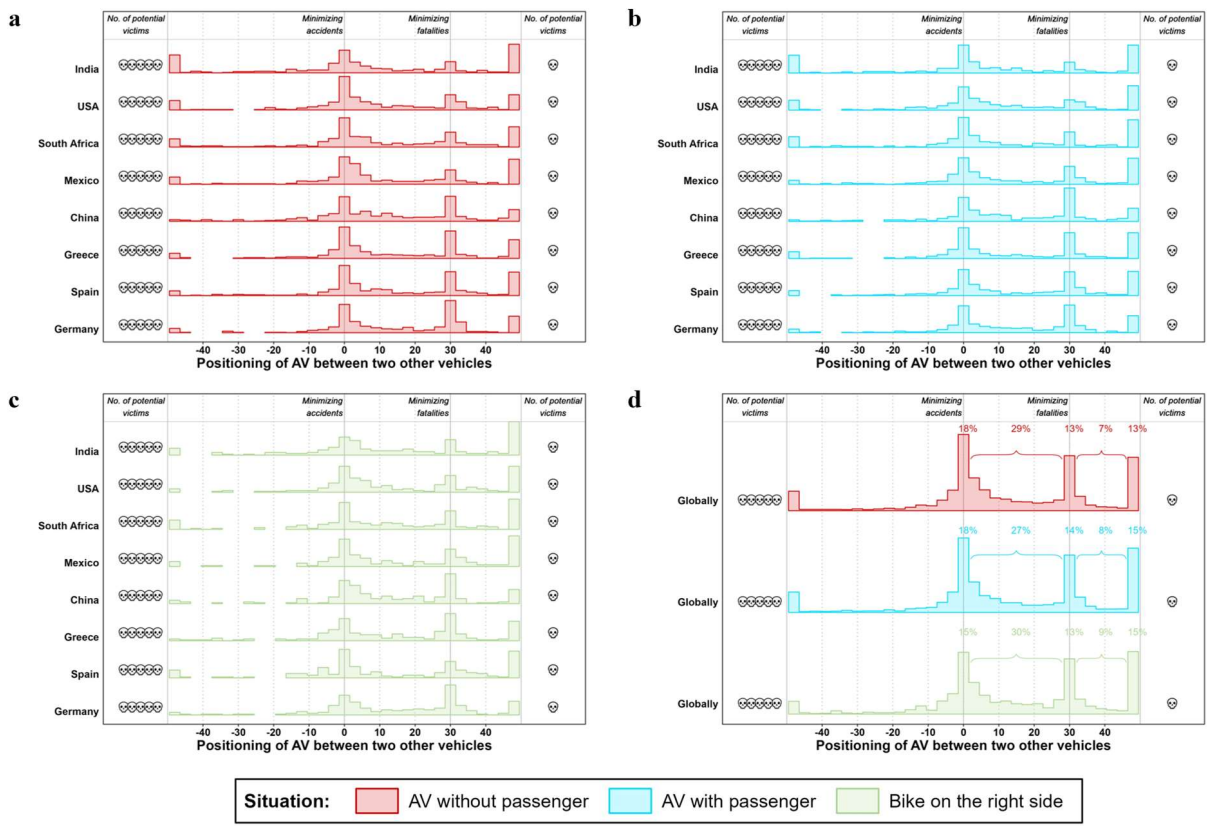

**Figure 5. Distributions of driving positions of the AV per country and globally.**
The figure shows four ridgeline plots with the positioning distributions of the (yellow) AV per country (Panels a-c) and globally when the data from all countries are pooled (Panel d). Each bar in the histograms comprises three adjacent AV positions. Bars at *0*, for instance, include positionings of the AV exactly at the lane's center as well as one deviation from the center to the left and to the right. Histograms in red refer to traffic situations when there was a (blue) car on either side of the road and the (yellow) AV was empty. Histograms in blue refer to situations when the participants were part of the traffic situation as a passenger of the AV. Histograms in green refer to situations when one cyclist appeared on the right side of the road.

When looking at the distributions based on the pooled data from all countries in panel (d), it is first of all noticeable that there are hardly any differences between the traffic situations. No matter whether the participants are part of the traffic situation or not, and no matter whether the types of road users are the same or not, the distributions of AV positioning are almost identical. Particularly with regard to the social dilemma of AVs, it is again surprising that it does not have any effect on the distribution of AV positioning whether the participants themselves are passengers in the AV or not. In a risk context, the social dilemma of AVs does not seem to play a role. Furthermore, it is noticeable that the vast majority of participants deviates from a guiding principle of minimizing accident probability. There is a clear discrepancy between the targeted behavioral principles of AVs in the academic engineering literature and the expectations of the (international) public in this regard. More than 60% of participants think that the greater number of potential accident victims should receive more protection in road traffic by increasing safety distances to them, even if this increases the overall probability of accidents.

It is important to emphasize that in the entire range of positions between the two minima of the traffic statistics (i.e., between the lane's center and a deviation of *+30*) there was a trade-off between the number of accidents and the number of fatalities. This means that in this range it was not possible to reduce the number of accidents without increasing the number of fatalities and vice versa. Therefore, from a social perspective, there is no clear superior AV position in this range if both objectives play a role. Outside of this range, the traffic statistics worsen in both dimensions, the number of accidents and the number of fatalities. From a social perspective, AV positions outside of this range should therefore be rejected due to Pareto-impairments. On an individual level, however, these positionings can of course still be rational. For example, if I will predominantly travel alone in my AV, it may be advantageous for me if fully occupied vehicles will carry a greater accident risk. AV positionings outside the socially preferred range are therefore not necessarily an expression of a mistake or misconception, even if we would like to see them that way from a social perspective.

## Discussion

The morally preferred distribution of risks between road users is strikingly similar between the eight different countries from diverse cultural regions of the world. This holds true for all three major findings of our study. These are, first, that most people deviate considerably from the guiding principle of accident avoidance and take an accident's potential severity into account when adjusting the AV's position within a lane. At the national level, the weighting of accident probability and severity was almost the same in all countries. Second, the social dilemma of AVs that was observed in deterministic contexts with unavoidable accidents evaporates in stochastic contexts in which accidents are possible but unlikely. Third, in none of the eight countries do risk preferences differ between road user types when controlling for varying vulnerability. Cyclists do not receive a bonus in the risk management of AVs compared to car drivers that goes beyond their higher vulnerability.

On an implicational level, our results suggest that a moral consensus on the risk ethics of AVs might be achievable. In fact, it may be more easily achievable than a moral consensus on the ethics of crashing in unavoidable accident scenarios where differences in moral attitudes could be identified between culturally diverse clusters (Awad et al., 2018). Especially the consistent mitigation of the social dilemma of AVs in stochastic contexts that we observe in our study seems encouraging when it comes to the global acceptance of AVs. In a globalized market, the prospect of AVs that do not have to fundamentally adjust their operating principles when crossing international borders to adapt to different risk preferences might help to realize economies of scale.

The advent of AVs offers the chance for a more deliberate management of traffic risks. Eliciting road users' preferred risk distributions allows to democratize the ethics of AVs by explicitly implementing people's moral preferences within the limits of technological feasibility. The desired behavior of AVs is first and foremost a political question on a societal scale (Himmelreich, 2018). Its technical feasibility is then a question for engineering. A lack of feasibility with today's technological means does not preclude feasibility with future technological means. The achievability of socially desirable solutions can, for instance, be steered by funding calls.

Beyond the specific functionality of AVs, however, the elicitation of road users' risk preferences might also help to democratize the risk ethics of the entire traffic system. Trade-offs are factually resolved in the specific design of our traffic infrastructure. In line with Vision Zero concepts in road traffic (Belin,

Tillgren & Vedung, 2012), for example, there is an increasing tendency to regulate junctions via roundabouts instead of traffic lights. The aim is to reduce the number of traffic fatalities and resolve the probability-severity trade-off discussed here at the expense of more crashes for the sake of less severe ones. The risk preferences revealed in our study are in line with such endeavors. Until autonomous driving matches the risk preferences of the public, it could be restricted to small and light vehicles to reduce their hazardousness to other road users in general (see, e.g., Tyndall, 2021; Anderson & Auffhammer, 2014). Similarly, traffic systems could be designed in such a way that road users can decide for themselves whether they want to participate in road traffic with AVs, especially as long as public expectations do not match the actual risk management of AVs. This could be achieved, for example, by creating AV-only lanes so that road users have the opportunity to opt into autonomous driving or by creating human-drivers-only lanes to allow users to opt out of traffic with AVs.

Our study is subject to limitations of which only two should be mentioned here. In our identification of a moral consensus, we relied on data from countries that were interpreted as representative for their cultural cluster as defined by the Inglehart-Welzel map. Therefore, it cannot be excluded that not selected countries exhibit risk distributions very different from the global consensus reported in our study. Moreover, we used only a limited amount of traffic situations in our study. Specifically, we focused on an imbalance of one versus five potential accident victims to check whether participants would systematically deviate from a lane's center when positioning the AV. Adding more combinations of the number of potential casualties to the analysis would have allowed to investigate the monotonicity of participants' positioning behavior. Similarly, a more nuanced study of positioning behavior would have been possible if more types of road users besides cars and cyclists had been added.

## Methods

The online experiment was conducted in December 2023 with the help of the survey service provider *Horizoom* (https://www.horizoom.de/). *Horizoom* was responsible for the management of the recruitment and payment of the study participants as well as the execution of representative quotas according to age (from 18 years) and gender in all countries of the study. The software for the experiment was programmed by us using oTree (Chen, Schonger & Wickens, 2016) and made available to the participants on Heroku's web servers (https://www.heroku.com/). The experiment received approval from the Institutional Review Board of the German Association for Experimental Economic Research (https://www.gfew.de/) and was conducted in accordance with the guidelines of the Declaration of Helsinki. In addition, we obtained informed consent from all participants prior to the start of the experiment and notified them that they could withdraw from the study at any time without consequences. The experiment was preregistered with all details, including the number of treatments, the number of participants and the planned data analysis at *AsPredicted* (https://aspredicted.org/) on November 27, 2023. The pre-registration of the experiment can be accessed using the study's registration number *152691* as well as any of the authors last name at the following link: https://aspredicted.org/lookup.php/.

The study was conducted in eight countries. These countries were selected on the basis of the Inglehart-Welzel Cultural Map of the World of 2023 (https://www.worldvaluessurvey.org/) by choosing one country from each of the eight country clusters for the study. These were (*in alphabetical order*) China (Confucian), Germany (Protestant Europe), Greece (Orthodox Europe), India (African-Islamic), Mexico (Latin America), South Africa (West & South Asia), Spain (Catholic Europe) and USA (English-Speaking). The experiment was conducted in a total of five languages. These were English (in the USA,

India, South Africa), Spanish (in Spain, Mexico), German (in Germany), Greek (in Greece) and Mandarin (in China). In total, 10,976 people completed the experiment, 1,372 in each country. On average, the participants were 39 years old, 48% of the participants were male, 52% female and 90% of the participants had a driving license. *Table 4* shows the age, gender and possession of a driving license of the participants in each country. With the help of the survey service provider, we were able to achieve a broad and balanced sample in all countries according to the participants' age and gender.

The online experiment had the following procedure (screenshots of the entire experiment can be found in the *Supplementary Material*). After the participants had given their informed consent to participate in the study (see Screen 1 in the *Supplementary Material*), the traffic situation, their task and the graphical interface were described to them (see Screens 2a and 2b in the *Supplementary Material*). A practice task further familiarized the participants with the redistribution of risks in an exemplary traffic situation (see *Figure 6*). We clearly emphasized if the participants themselves were part of the traffic situation (see *Figure 6* as well as Screen 2b in the *Supplementary Material*). Subsequently, the participants had to answer two comprehension questions (see Screen 3 in the *Supplementary Material*). These questions were easy to answer if the participants had read the instructions and tried out the practice task (participants could not proceed to Screen 3 without trying out the practice task at least once). As preregistered, participants could only take part in the experiment if both comprehension questions were answered correctly. All other participants were excluded from the experiment.

|  | Age |  |  | Gender |  |  | Driver's license (= Yes) | Bat-and-ball (= correct) |
|---|---|---|---|---|---|---|---|---|
|  | Mean (S.D.) | Min | Max | Male | Female | Other |  |  |
| Germany | 44.7 (14.1) | 18 | 69 | 43.9% | 55.8% | 0.3% | 89.7% | 20.4% |
| Spain | 43.6 (13.4) | 18 | 78 | 49.7% | 50.2% | 0.1% | 92.1% | 10.6% |
| Greece | 42.7 (13.0) | 19 | 93 | 50.8% | 49.2% | --- | 90.8% | 14.7% |
| China | 34.3 (8.1) | 18 | 65 | 46.3% | 53.6% | 0.1% | 95.6% | 43.4% |
| Mexico | 37.1 (11.9) | 18 | 73 | 44.6% | 55.3% | 0.1% | 82.2% | 5.2% |
| South Africa | 35.8 (12.3) | 18 | 70 | 45.8% | 54.2% | --- | 89.5% | 3.6% |
| USA | 42.3 (14.3) | 18 | 95 | 47.2% | 52.4% | 0.4% | 90.7% | 10.2% |
| India | 34.5 (11.5) | 18 | 74 | 52.6% | 47.4% | --- | 88.7% | 11.5% |

**Table 4. Demographics of participants.**
Some demographic characteristics of the country samples based on self-reports of participants in post-experimental questionnaire.

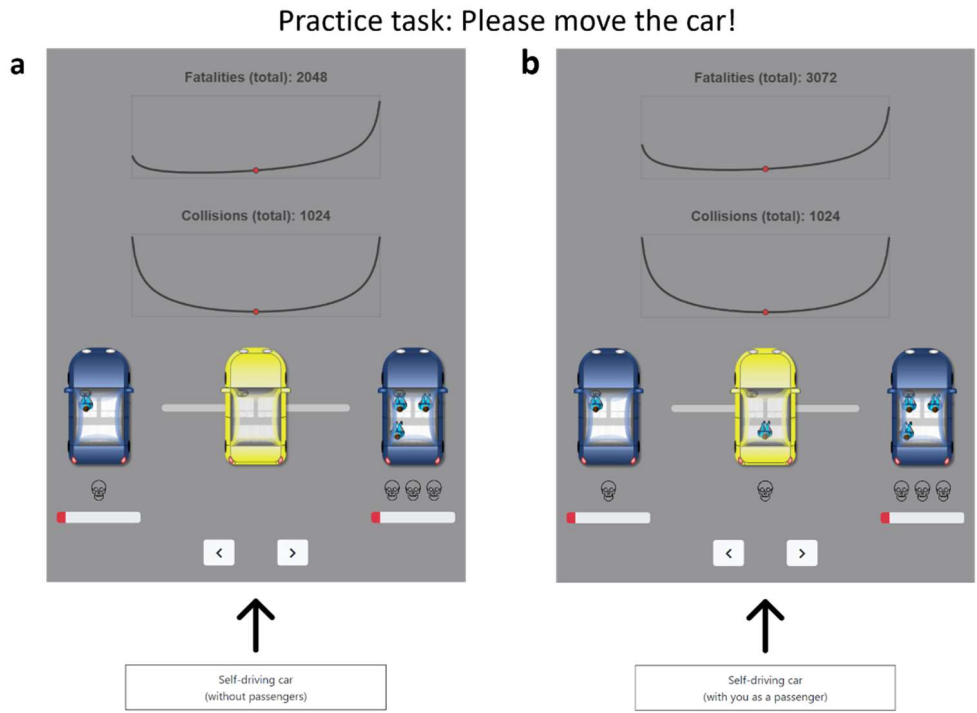

**Figure 6. Practice task on introductory screen with description of the situation.**
The figure shows two versions of the practice task, depending on whether the participants were part of the traffic situation as passengers in the (yellow) AV or not. The screen contained further information (see Screens 2a and 2b in the *Supplementary Material*) and participants could only proceed to the next screen if they moved the (yellow) AV at least once.

If the participants answered both comprehension questions correctly, they were first informed of this (see Screen 4 in the *Supplementary Material*). Afterwards, we confronted them with the traffic situation in which they were asked to position the AV between the two other vehicles at their discretion (see Screens 5a to 5c in the *Supplementary Material*). Participants could adjust the driving position of the AV in 99 increments. The initial position of the (yellow) AV was chosen at random for each participant. All participants were confronted with only one traffic situation. This situation was randomly selected for each participant from a total of seven possible situations. In all seven situations, a (yellow) AV was driving between two other vehicles. In six of the seven situations, the vehicles on both sides of the road were cars; in one other situation, there was a car on the left and a cyclist on the right side of the road. In three traffic situations, the (yellow) AV was driving without passengers; in the other four situations, the AV was occupied by the respective participant as a single passenger. If the (yellow) AV was driving without passengers, the traffic situations always contrasted one accident victim on one side of the road with five accident victims on the other side. If the (yellow) AV was driving with the respective participant as a single passenger, two of the four traffic situations contrasted one accident victim on one with five accident victims on the other side of the road; the other two situations contrasted four accident victims on one with zero accident victims on the other side of the road. *Table 5* provides an overview of all traffic situations and shows the respective number of participants per country for each situation.

As in Krügel & Uhl (2024), we visualized the accident probability with the left or right vehicle using a red bar below the corresponding vehicle. In addition, we presented to the participants two traffic statistics above the (yellow) AV: the estimated total number of (*i*) collisions and (*ii*) fatalities in one million such traffic situations. The order of both statistics was set at random for each participant. The respective numbers of the statistics changed depending on the positioning of the (yellow) AV. A red dot marked the current point on the corresponding graph of the function. For the accident probability, we implemented an exponential relationship between collision probability and safety distances between two adjacent vehicles. The overall probability of an accident was lowest when the (yellow) AV was driving exactly in the lane's middle. This probability increased exponentially with deviations from the middle driving position. The smaller the distance to one of the two other vehicles, the greater the probability of an accident with this vehicle (see also Krügel & Uhl, 2024). To calculate the total number of fatalities, the accident probability of the AV with the vehicle on the left and right side of the road was multiplied by the respective number of accident victims and summed up. For the sake of simplicity, this calculation was based on the assumption that all accidents are fatal for all persons involved in the collision (including the passenger of the (yellow) AV).

The statistics presented to the participants were purely hypothetical, but not arbitrary. When generating these statistics, it was important to us that every possible driving position of the (yellow) AV in a traffic situation was associated with a unique combination of the number of collisions and fatalities. In addition, it was essential that there was a trade-off between collision avoidance and minimization of fatalities in each traffic situation. The minimum of each objective was therefore associated with a different driving position. The participants were free to choose their preferred positioning of the AV in this trade-off according to their moral convictions. *Table 6* provides an overview of means and standard deviations of participants' positioning of the (yellow) AV per country in all traffic situations.

After positioning the (yellow) AV, the participants answered a short post-experimental questionnaire in which we asked about their age, gender and whether they had a driver's license (see Screen 6 in the *Supplementary Material*). In addition, we confronted the participants with the "bat and ball problem" (Frederick, 2005) of the cognitive reflection test (see Screen 7 in the *Supplementary Material*). Afterwards, the online experiment was completed.

| Middle car (AV) | Without passenger | | | With passenger | | | | |
|---|---|---|---|---|---|---|---|---|
| Vehicle on the right side | Car | | Bike | Car | | Car | | |
| No. of victims on left vs. right side | 5 vs. 1 | 1 vs. 5 | 5 vs. 1 | 5 vs. 1 | 1 vs. 5 | 4 vs. 0 | 0 vs. 4 | Total |
| Germany | 199 | 208 | 189 | 198 | 196 | 199 | 183 | 1,372 |
| Spain | 190 | 208 | 188 | 197 | 196 | 193 | 200 | 1,372 |
| Greece | 202 | 189 | 191 | 193 | 199 | 204 | 194 | 1,372 |
| China | 196 | 190 | 196 | 193 | 195 | 203 | 199 | 1,372 |
| Mexico | 202 | 191 | 196 | 206 | 190 | 196 | 191 | 1,372 |
| South Africa | 205 | 194 | 193 | 196 | 199 | 196 | 189 | 1,372 |
| USA | 199 | 201 | 194 | 193 | 202 | 183 | 200 | 1,372 |
| India | 186 | 206 | 189 | 201 | 201 | 194 | 195 | 1,372 |

**Table 5. Overview of experimental conditions.**
Number of observations per experimental condition and country.

| Middle car (AV) | Without passenger | | | With passenger | | | | |
|---|---|---|---|---|---|---|---|---|
| Vehicle on the right side | Car | | Bike | Car | | Car | | |
| No. of victims on left vs. right side | 5 vs. 1 | 1 vs. 5 | 5 vs. 1 | 5 vs. 1 | 1 vs. 5 | 4 vs. 0 | 0 vs. 4 | Total |
| Germany | 15.1 (21.6) | -16.2 (21.0) | 15.3 (20.8) | 12.6 (21.6) | -9.5 (23.1) | 15.4 (23.4) | -15.1 (19.4) | 2.5 (25.8) |
| Spain | 18.4 (23.0) | -12.3 (23.3) | 13.6 (25.8) | 9.7 (21.8) | -6.2 (22.4) | 19.8 (21.4) | -13.0 (24.9) | 4.0 (26.8) |
| Greece | 12.8 (21.4) | -14.2 (24.0) | 14.7 (22.5) | 11.4 (20.2) | -6.7 (22.2) | 17.9 (23.6) | -11.8 (25.0) | 3.6 (26.0) |
| China | 13.4 (22.6) | -12.3 (19.5) | 13.9 (20.8) | 12.0 (18.1) | -14.1 (18.7) | 16.5 (20.1) | -14.6 (21.3) | 2.2 (24.4) |
| Mexico | 12.4 (25.0) | -12.2 (23.2) | 16.6 (23.8) | 13.5 (23.0) | -8.9 (21.8) | 17.1 (22.9) | -12.6 (24.4) | 4.0 (26.8) |
| South Africa | 13.1 (24.9) | -7.5 (25.6) | 14.3 (25.9) | 9.8 (25.3) | -6.7 (22.7) | 15.9 (26.1) | -6.4 (27.6) | 4.8 (27.4) |
| USA | 12.2 (23.1) | -7.3 (24.6) | 15.1 (22.6) | 9.7 (20.5) | -4.2 (23.5) | 13.8 (25.5) | -9.1 (26.6) | 4.1 (25.8) |
| India | 12.2 (28.1) | -2.7 (31.2) | 14.7 (26.2) | 12.0 (28.8) | 2.8 (27.8) | 14.2 (27.5) | -1.9 (31.1) | 7.2 (29.6) |

**Table 6. Average positioning of AV per country in all traffic situations.**
The table shows means and, in parentheses, standard deviations of participants' positioning of the (yellow) AV per country in all traffic situations. Positive mean values indicate an average deviation from the lane's center to the right; negative mean values indicate an average deviation from the lane's center to the left.

# Supplementary Material

Below you will find all the screens of the survey in English. Screens in other languages (Spanish, German, Greek or Mandarin) are available upon request.

**Fig. S1. Screen 1: Informed consent.**

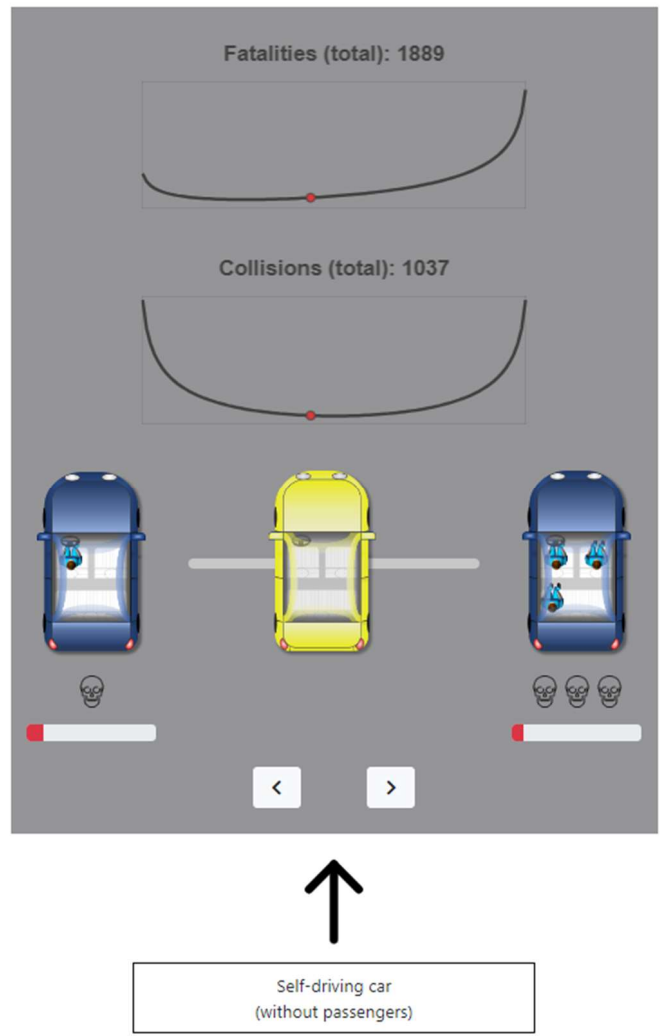

**Fig. S2. Screen 2a: Description of the traffic situation, the task and graphical interface (treatments *AV without passenger* and *Bike on the right side*).**

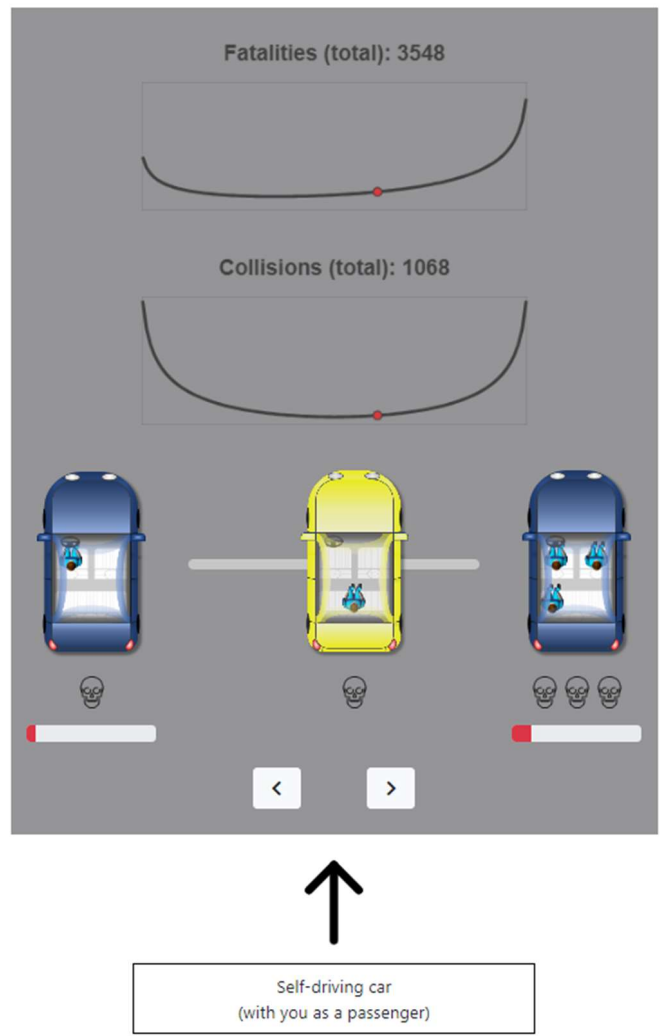

**Fig. S3. Screen 2b: Description of the traffic situation, the task and graphical interface (treatment *AV with passenger*).**

**Comprehension questions**

What is your task?
○ I am to position the yellow self-driving car as I think it is right.
○ I am to position the yellow self-driving car in the way that others think is right.

If you move the yellow self-driving car, the estimated total number of fatalities and collisions in one million such traffic situations will also change.
○ Yes, that's true.
○ No, that's not true.

[Next]

**Fig. S4. Screen 3: Both control questions (all treatments).**

You have answered everything correctly!
**Please click on the button to start the survey.**

[Next]

**Fig. S5. Screen 4: Survey continues if both control questions were answered correctly.**

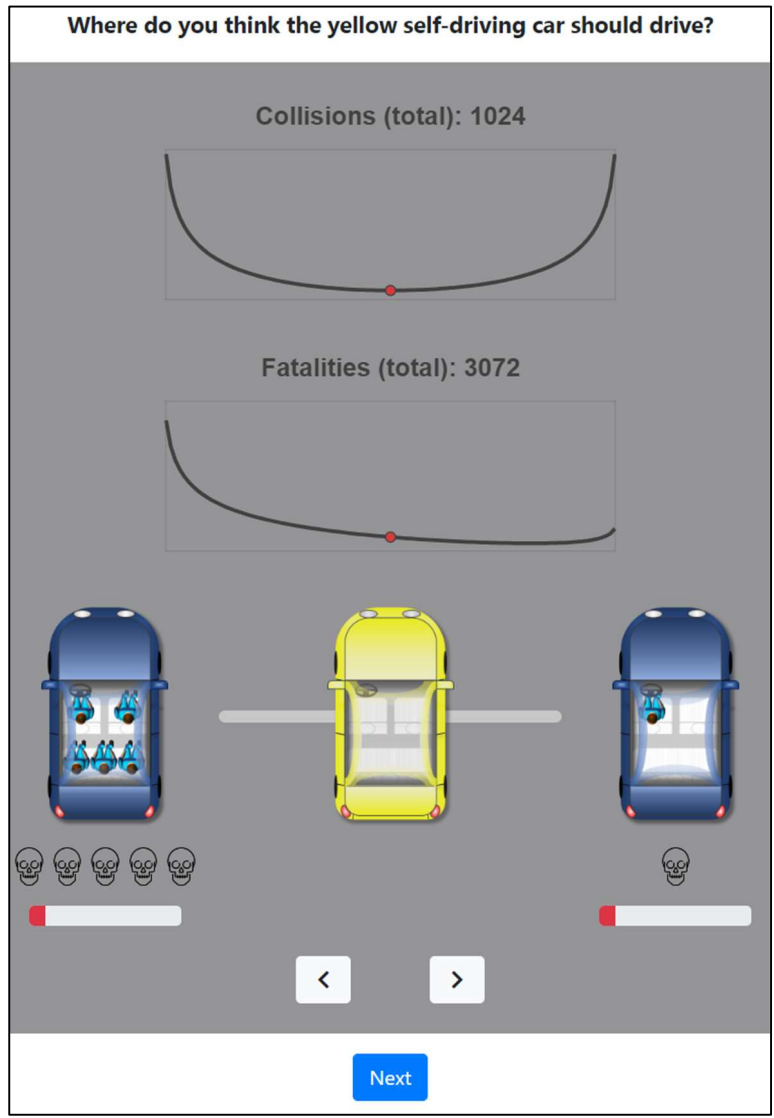

**Fig. S6. Screen 5a: Decision screen (treatment *AV without passenger*).**

Participants were able to drag the (yellow) AV back and forth between both other vehicles in 99 increments. The initial position of the AV was chosen at random for each participant. The red bar below each vehicle visualized the collision probability as a function of the distance of the AV. The closer the AV to a vehicle, the greater [smaller] the probability of a collision with that [the opposite] vehicle. The red bar below each vehicle increased [decreased] accordingly. We implemented an exponential relationship between the distance of the AV and the collision probability with a vehicle according to the following function, where the middle driving position of the AV between the two other vehicles minimized the overall accident probability:

$$P(collision) = \left(\frac{1}{\sqrt[8]{x}} - 0.562\right) \times 10^{-2}$$

with $x \in \{1, 2, 3, \dots, 99\}$ representing the distance of the AV to the respective vehicle. The two graphs above the (yellow) AV showed the estimated total number of collisions and fatalities in one million such traffic situations. To calculate the total number of fatalities, the accident probability of the AV with the vehicle on the left and right side of the road was multiplied by the respective number of accident victims and summed up. The order of both statistics was set at random for each participant. The respective numbers of the statistics changed depending on the positioning of the (yellow) AV and a red dot marked the current point on the corresponding graph of the function. Each driving position of the AV was associated with a unique combination of the number of collisions and fatalities.

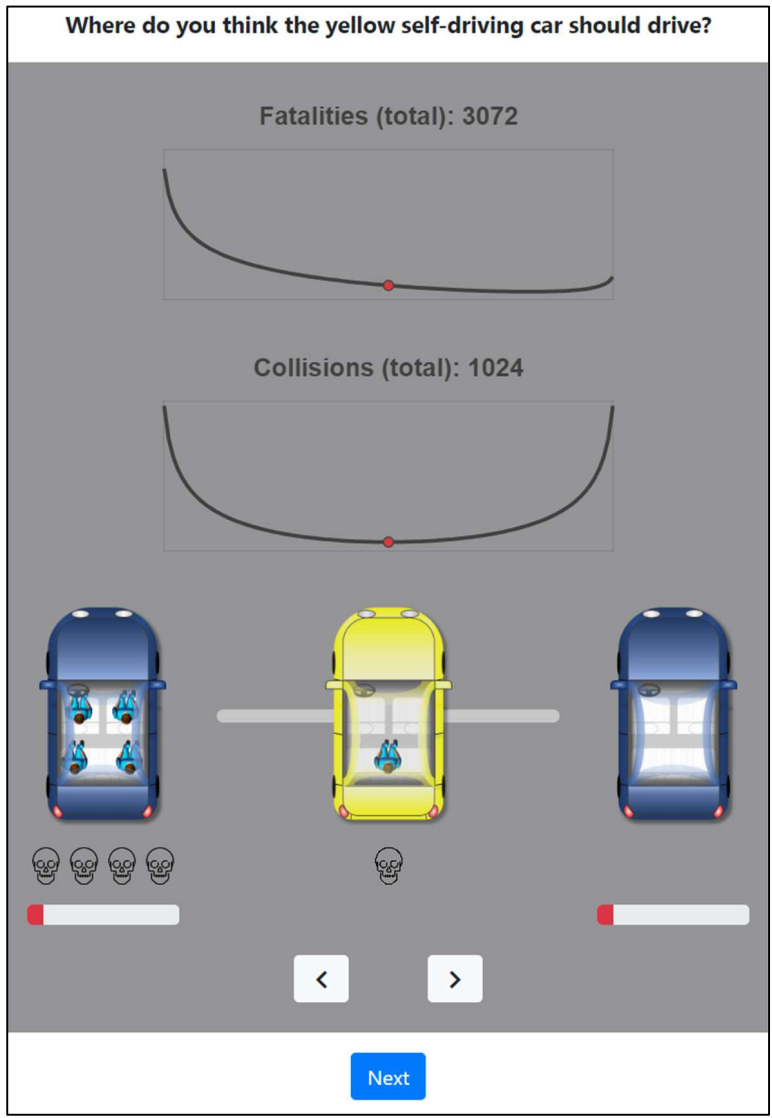

**Fig. S7. Screen 5b: Decision screen (treatment *AV with passenger*).**

(*Also see the notes to Fig. S6.*)

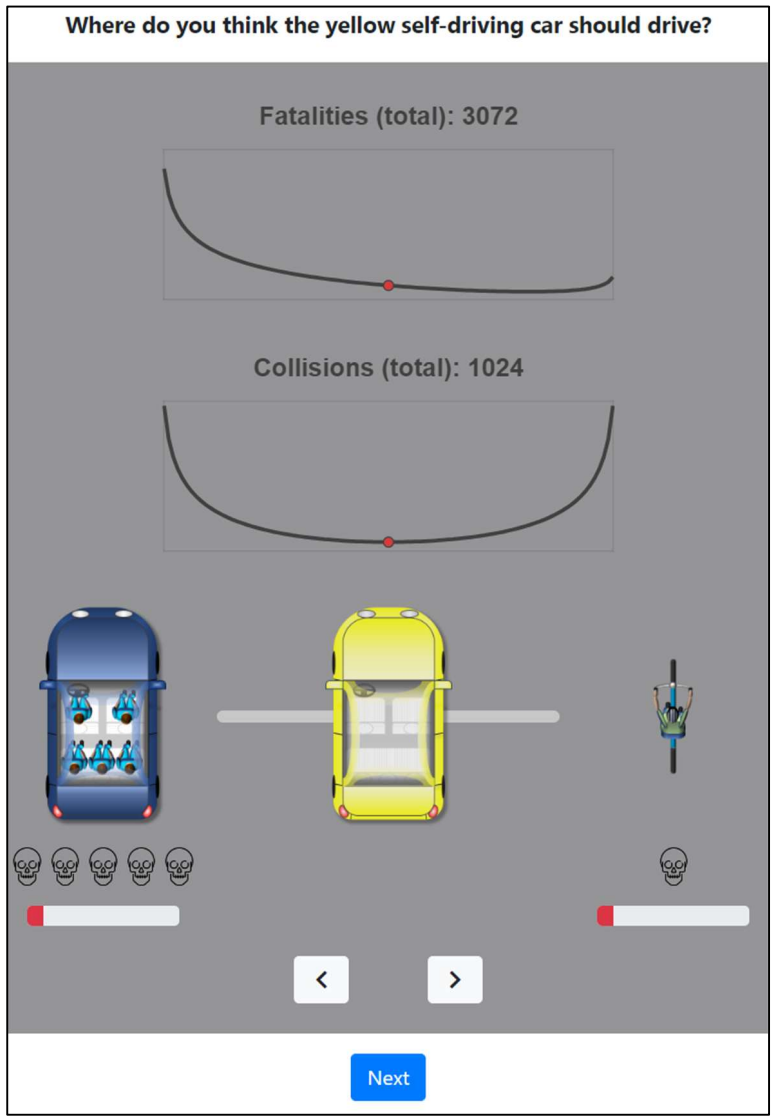

**Fig. S8. Screen 5c: Decision screen (treatment *Bike on the right side*).**

(*Also see the notes to Fig. S6.*)

**Fig. S9. Screen 6: Demographic and personal characteristics.**

**Fig. S10. Screen 7: One question of the cognitive reflection test.**

**Fig. S11. Screen 8: End of survey.**